\documentclass[prl,twocolumn,amsfonts,amsmath,amssymb,floatfix]{revtex4}
\usepackage[latin1]{inputenc}
\usepackage{latexsym}
\usepackage{graphicx}
\usepackage{verbatim}
\newlength{\tabsep}
\newlength{\bb}
\newlength{\dbb}
\newcommand{\mb}{\mathbf}

\begin{document}

\title{Outer-totalistic cellular automata on graphs}

\author{Carsten \surname{Marr}}
\email{carsten.marr@helmholtz-muenchen.de} \affiliation{Institute for
  Bioinformatics and Systems Biology, Helmholtz Zentrum München -
  German Research Center for Environmental Health, D-85764 Neuherberg,
  Germany} 
\author{Marc-Thorsten \surname{H\"utt} }
\email{m.huett@jacobs-university.de} \affiliation{Computational
  Systems Biology, School of Engineering and Science, Jacobs
  University Bremen, D-28759 Bremen, Germany}

\date{\today}

\begin{abstract} 
  We present an intuitive formalism for implementing cellular automata
  on arbitrary topologies. By that means, we identify a symmetry
  operation in the class of elementary cellular automata. Moreover, we
  determine the subset of topologically sensitive elementary cellular
  automata and find that the overall number of complex patterns
  decreases under increasing neighborhood size in regular
  graphs. As exemplary applications, we apply the formalism to complex
  networks and compare the potential of scale-free graphs and
  metabolic networks to generate complex dynamics.
\end{abstract}

\maketitle

\textit{Introduction}---Cellular automata (CA) on graphs in principle
provide the possibility to monitor systematic changes of dynamics
under variation of network topology. In practice, however,
unambiguously studying the relation between topology and dynamics with
CA is conceptually difficult, since changes in topology inevitably
induce changes in the rule space. Proposed by \citet{neumann63} as a
model system for biological self-reproduction, a surge of research
activity from the 80's onwards \citep{wolfram83} established them as
the standard tool of complex systems theory and spatio-temporal
pattern formation \cite{deutsch05} on regular grids. Another discrete
and binary modeling approach for complex biological systems are random
Boolean networks (RBNs), introduced by \citet{kauffman69}. While the
CA framework introduces one rule for all regularly ordered cells with
bi-directional links, the original RBNs consist of randomly and
directionally linked nodes with individual rules.  
Here, we present a formalism that generalizes CA to
arbitrary architectures. It allows (i) the establishment of a general
correspondence between CA and isotropic RBNs and (ii) the comparison
of the potential of different topologies to generate complex
dynamics. As applications we examine the topological sensitivity of
elementary CA, monitor the number of complex rules of CA under
increasing neighborhood size, and compare the dynamic potential of
scale-free graphs and representations of metabolism as substrate
graphs.

\textit{The formalism}---Within the CA framework, the discrete
(binary) state $x_i \in \Sigma = \{0,1\}$ of a node $i$ at time $t+1$
solely depends on its own state and the states of its $d$ neighboring
nodes at time $t$. All cells are updated synchronously by the same,
time-independent rule $f: \Sigma^{d+1} \rightarrow \Sigma$. To
implement CA on a directed or undirected graph $G$, we have to
account for different neighborhood sizes $d_i$ due to the heterogeneous
connectivity and thus, in general, to allow for individual rules
$f_i$. Our strategy instead is to impose constraints on the rule
space, motivated by simple physical requirements, in order to obtain a
set of discrete rules, implementable on arbitrary topologies:
\begin{itemize}
\item Homogeneity $f_i=f \; \forall \; i$, i.e.~the same rule
  applies to all nodes in the graph.
\item Isotropy $f=f(x_i, \rho_i)$, i.e.~rules may not depend on the order
  of neighboring states and are thus functions of the density of
  neighboring states, $\rho_i(t) = \frac{1}{d_i} \sum_j A_{ij}
  x_j(t)$. Here, $G$ is represented by the adjacency matrix
  $\mb A$: If a link connects node $j$ to node $i$, $A_{ij} = 1$, and we
  call $j$ an input node of $i$.  The number of all input nodes is
  called the in-degree of node $i$, $d_i= \sum_j A_{ij}$.
\item Functional simplicity, i.e.~the rule $f$ is a piecewise constant
  function of the density $\rho_i$. 
\end{itemize}

\textit{Elementary Cellular Automata}---The simplest CA, termed
elementary CA (ECA) \cite{wolfram83}, are defined on a one-dimensional
grid with minimal neighborhood size, $d=2$, and a binary state space,
$\Sigma = \{0,1\}$. The $2^3 = 8$ different neighborhood
configurations $x_{i-1},x_i,x_{i+1}$ result in $2^{2^3}=256$ possible
rules. In this set, $2^6=64$ rules fulfill the conditions mentioned
above and depend only on the state $x_i$ ($0$ or $1$) and on the
density $\rho_i$ of neighboring states (0, 1/2, or 1). These 64 rules
are called outer-totalistic \cite{wolfram83} and are now parametrized
with the rule parameter set ($\alpha$, $\beta$, $\gamma$):
 \begin{equation}
  x_i(t+1) =
  \begin{cases}
    \alpha \,,    &  \rho_i = 0 \\
    \beta \,,    &  \rho_i = 1/2\\
    \gamma \,,    &  \rho_i = 1
  \end{cases}
\label{eq:2}
\end{equation}
We distinguish the following cases for the rule parameters
$\alpha,\beta,\gamma$: The state $x_i(t+1)$ may be $0$ or $1$
independently of the state $x_i(t)$ itself, or it may remain unchanged
($+$) or be flipped ($-$), $\alpha,\beta,\gamma \in
\{0,1,+,-\}$. The frequently used majority rule
\cite{crutchfield95_proceedings-of-the-n,moreira04,amaral04,nochomovitz06},
for example, where a node $i$ is mapped onto $0/1$ if the density
$\rho_i$ is below$/$above $0.5$, and stays in its state otherwise, is
described in our formalism by $(\alpha,\beta,\gamma) = (0,+,1)$.
For $\alpha,\beta,\gamma \in \{0,1\}$, the corresponding CA rules
are called totalistic \cite{wolfram83}, since $x_i (t+1)$ depends
exclusively on the density $\rho_i$ of the input states. Only these
rules have strict RBN rule equivalents (see Table \ref{tab:gca}).

Aside from the initial system state $\mb{x}(0) := (x_1, x_2, \dots,
x_N)$ at $t=0$, the patterns of rule $(0,0,0)$ and rule $(1,1,1)$ are
perfectly symmetric under the action of the operator $\mathcal{T}: \xi
\mapsto 1-\xi, \; \xi \in \{0,1\}$. The operator $\mathcal{T}$
exchanges all 0s and 1s in an array of elements, which can be both a
pattern consisting of 0's and 1's or a set of rule parameters. Note
that the elements $\{+,-\}$ remain unaffected under the action of
$\mathcal{T}$. Generally, the symmetric rule to
$(\alpha,\beta,\gamma)$ is rule $\mathcal{T} (\gamma,\beta,\alpha)$.
The patterns emerging from the action of a rule onto an initial state,
written as $(\alpha, \beta, \gamma) \cdot \mb{x}(0)$, are identical to
the inverted patterns emerging from the inverted initial state
$\mathcal{T} \mb{x}(0)$ due to $\mathcal{T} (\gamma,\beta,\alpha)$:
$(\alpha,\beta,\gamma) \cdot \mb{x}(0) = \mathcal{T}
(\gamma,\beta,\alpha) \cdot \mathcal{T} \mb{x}(0)$. Explicitly, the
symmetric rule to $(0,1,+)$, corresponding to the ECA with rule number
218 \cite{wolfram83}, is $(+,0,1)$ with ECA rule number 164 (see Table
\ref{tab:gca} for more examples). Some rules, like the majority rule
$(0,+,1)$, are self-symmetric. After elimination of all symmetric
counterparts, 34 different ECA rules remain.

\begin{figure}[b] 
\begin{center}
    \includegraphics[width=0.8\columnwidth]{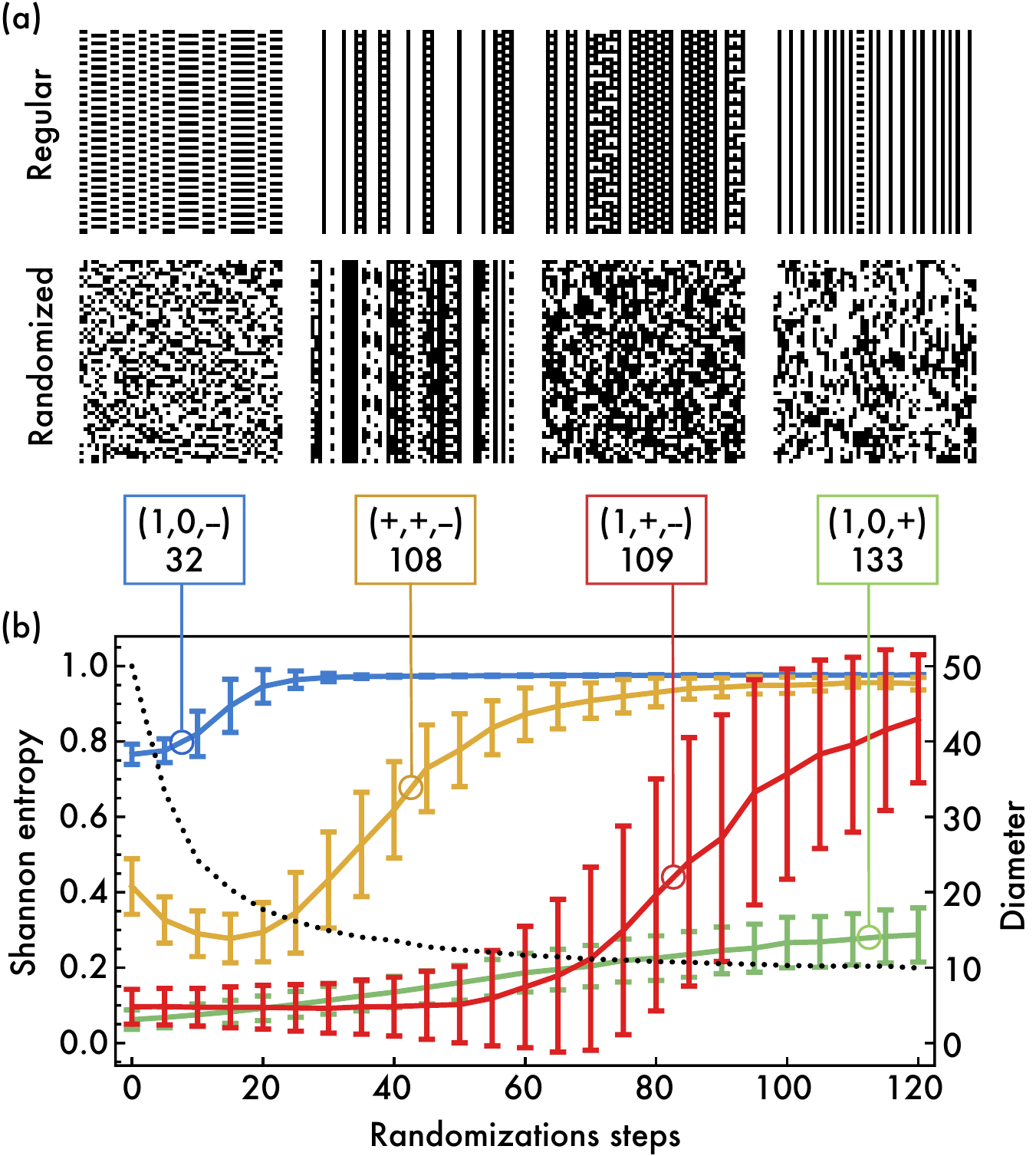}
    \caption{Rules $(1,0, -)$, $(+, +, -)$, $(1, +, -)$, and $(1, 0,
      +)$ (in ECA terms 37, 108, 109, and 133), on graphs with $N=100$
      nodes and $d=2$. (a) The upper row shows patterns from a regular
      ECA architecture, the lower row from randomized counterparts.
      Time runs downwards. (b) The Shannon entropy $S$ changes
      differently for every one of the four rules (colored lines)
      under randomization and is not trivially correlated with the
      variation of the diameter of the graph (dotted black line), as a
      prominent topological observable. The results are qualitatively
      independent of the network size.}
  \label{patterns}
\end{center}
\end{figure}


Which of these 34 rules are topologically sensitive? That is, which
lead to patterns of considerably different complexity when implemented
on the regular ECA grid and the RBN architecture?  Wolfram classified
CA heuristically according to the complexity of the emerging patterns
into the four Wolfram classes \cite{wolfram84}. On graphs, this
classification inevitably fails because of a lacking natural node
order. Instead, we apply two entropy-like measures, the Shannon
entropy $S$ and the word entropy $W$, which we have previously shown
to provide a feasible framework for the quantification of pattern
complexity
\cite{marr06_physics-letters-a,marr05_physica-a,marr07_pre}.  The
Shannon entropy $S$ serves as a measure for the homogeneity of the
spatio-temporal pattern, by averaging over all nodes: $S = \frac{1}{N}
\sum_{i=1}^N - (p^0_i \log_2 p^0_i + p^1_i \log_2 p^1_i )$. The
probabilities $p^0_i$ and $p^1_i$ denote the ratios of 0's and 1's in
the time series of node $i$.  The word entropy $W$ serves as a
complexity measure beyond single time steps. It quantifies the
irregularity of a time series by counting the number of words,
i.e.~blocks of constant states confined by the respective different
state: $W = \frac{1}{N} \sum_{i=1}^N \left( -\sum_{l=1}^{t} p_i^l
  \log_2 p_i^l \right)$. The probability $p_i^l$ is the number of
words of length $l$ divided by the number of all words found in the
time series of node $i$. The maximal possible word length is given by
the length $t$ of the time series analyzed.

We compare $10^3$ random initial conditions on the regular
architecture with $10^3$ samples of a randomized graph, where the
number of incoming and outgoing links of every node is preserved and
kept to $d=2$, but the link architecture has been randomized
\cite{trusina04}. Notably, for a considerably large number of
randomization steps, we generate random regular graphs
\cite{wormald99_randreggraphs} rather than Poisson-distributed random graphs
\cite{erdos59}.

We classify rules as topologically sensitive, if the difference of the
mean entropies for regular and randomized architectures is beyond the
standard deviation of the difference. Out of the 34 rules, 14 rules
fulfill that condition for at least one observable, $S$ or $W$. These
can be divided into three groups: (1) For some rules, the ratio of
constant and oscillating nodes changes under randomization, but
complex or chaotic patterns never occurs. (2) Others exhibit chaotic
patterns on both topologies, but the amount of complexity varies. (3)
The most interesting rules are those, for which the change of topology
leads to a fundamental change in the complexity of the resulting
patterns, i.e., a change in the Wolfram class. These rules are $(1, 0,
-)$, $(+, +, -)$, $(1, +, -)$, and $(1, 0, +)$ (in ECA terms rules 37,
108, 109, and 133). Typical time evolutions of these four rules on
regular and random architectures are shown in Figure
\ref{patterns}(a). Moreover, these four rules react specifically to
topological changes. Figure \ref{patterns}(b) shows the Shannon
entropy against the number of randomization steps performed. While the
patterns of rules $(1,0,-)$ and $(+,+,-)$ change already when a small
number of shortcuts are introduced into the system, $(1,+,-)$ stays
constant in this regime but shows higher $S$ and large variations for
strongly disordered topologies. Finally, the Shannon entropy of the
patterns emerging from rule $(1,0,+)$ grows monotonously with the
randomization depth.


\begin{figure}[b]
\begin{center}
\includegraphics[width=0.8\columnwidth]{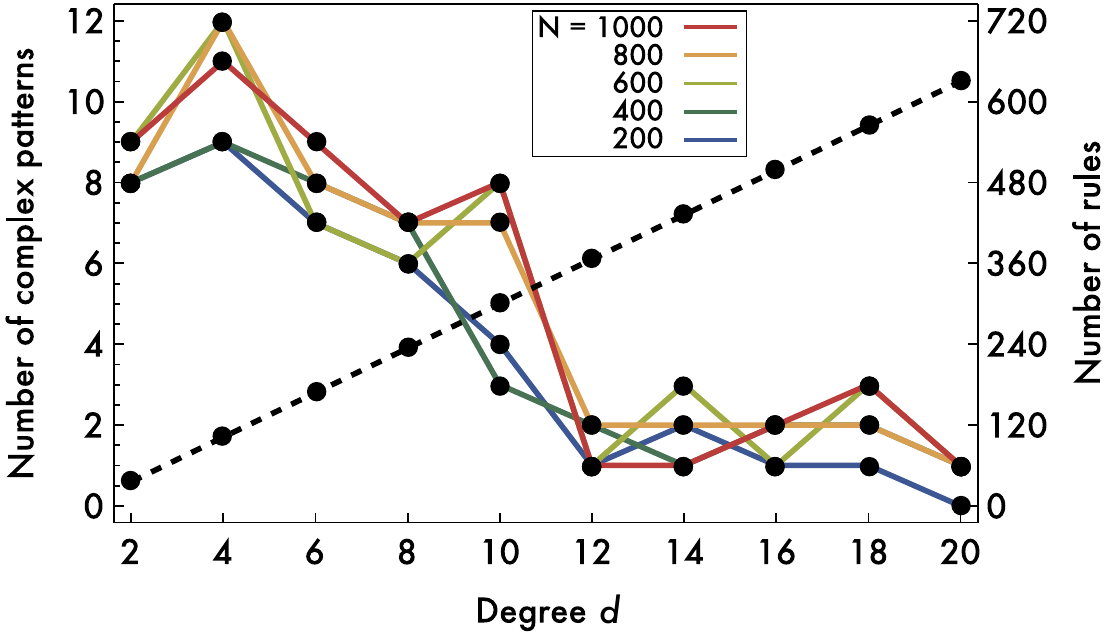}
\caption{Number of complex patterns (solid colored lines) and number
  of rules (dashed line) for regular graphs of different networks size
  against degree $d$.  The absolute number of possible rules increases
  due to the enlarged possible set of $\kappa$ values. Strikingly, the
  number of single threshold rules with complex patterns is maximal at
  $d=4$.}
  \label{neighbors}
\end{center}
\end{figure}

\textit{Regular graphs}---How much complexity is possible on regular
graphs?  With growing neighborhood size $d$, the number of possible
densities $\rho$ and therefore the number of possible rules increases.
For the sake of simplicity, we restrict our investigation to a binary
state space and to rules with a single threshold: $\kappa$:
\begin{equation} x_i(t+1) =
  \begin{cases}
    \alpha \,,    &  \rho_i < \kappa \\
    \beta \,,    &  \rho_i = \kappa \\
    \gamma \,, & \rho_i > \kappa
  \end{cases}
\label{eq:3}
\end{equation}
For networks with $d=12$, 11 different threshold parameters $\kappa
\in \{1/12, 2/12, \ldots, 11/12\}$ lead to 336 different rules, where
symmetric rules are considered only once. To estimate the number of
rules with complex (that is in our context: non-trivial) patterns, we
calculate $W$ for time evolutions. The word entropy is a feasible
complexity measure for individual time evolutions. It however fails to
disentangle periodic patterns (Wolfram class II) from complex (Wolfram
class IV) ones. We therefore supplement our classification with a
detrended fluctuation analysis (DFA) \cite{peng94}. This method
characterizes the time correlations of a signal with a single scaling
exponent, by calculating the variance of the signal from its trend in
a time window for different window sizes. The DFA exponent is the
slope of the mean variance against the window size and lies between
0.5 and 1.5 for white and Brownian noise, respectively. Applied to the
time evolution of the system's state density $\bar{\rho}(t) = \sum_i
x_i(t)$ as, e.g., in \cite{amaral04}, it can be used to discriminate
stationary and periodic patterns from complex ones.  We count patterns
as complex, if the DFA exponent is positive and the word entropy
$W>1$. However, the exact values of these thresholds do not alter our
results qualitatively.  As shown in Figure \ref{neighbors}, the number
of possible rules (dashed line) increases linearly with $d$, while a
maximum of complex patterns (full lines) occurs for $d=4$. The
striking overall reduction of complexity for neighborhood enlargement,
seen as the most dominant effect in Figure \ref{neighbors} can be
understood qualitatively from a homogeneity rationale: In the limit of
a fully connected graph, all nodes see nearly the same neighborhood
and thus follow the same dynamics (see \cite{marr05_physica-a} for a
more detailed explanation of a corresponding phenomenon).

\begin{figure}[b]
\begin{center}
\includegraphics[width=0.8\columnwidth]{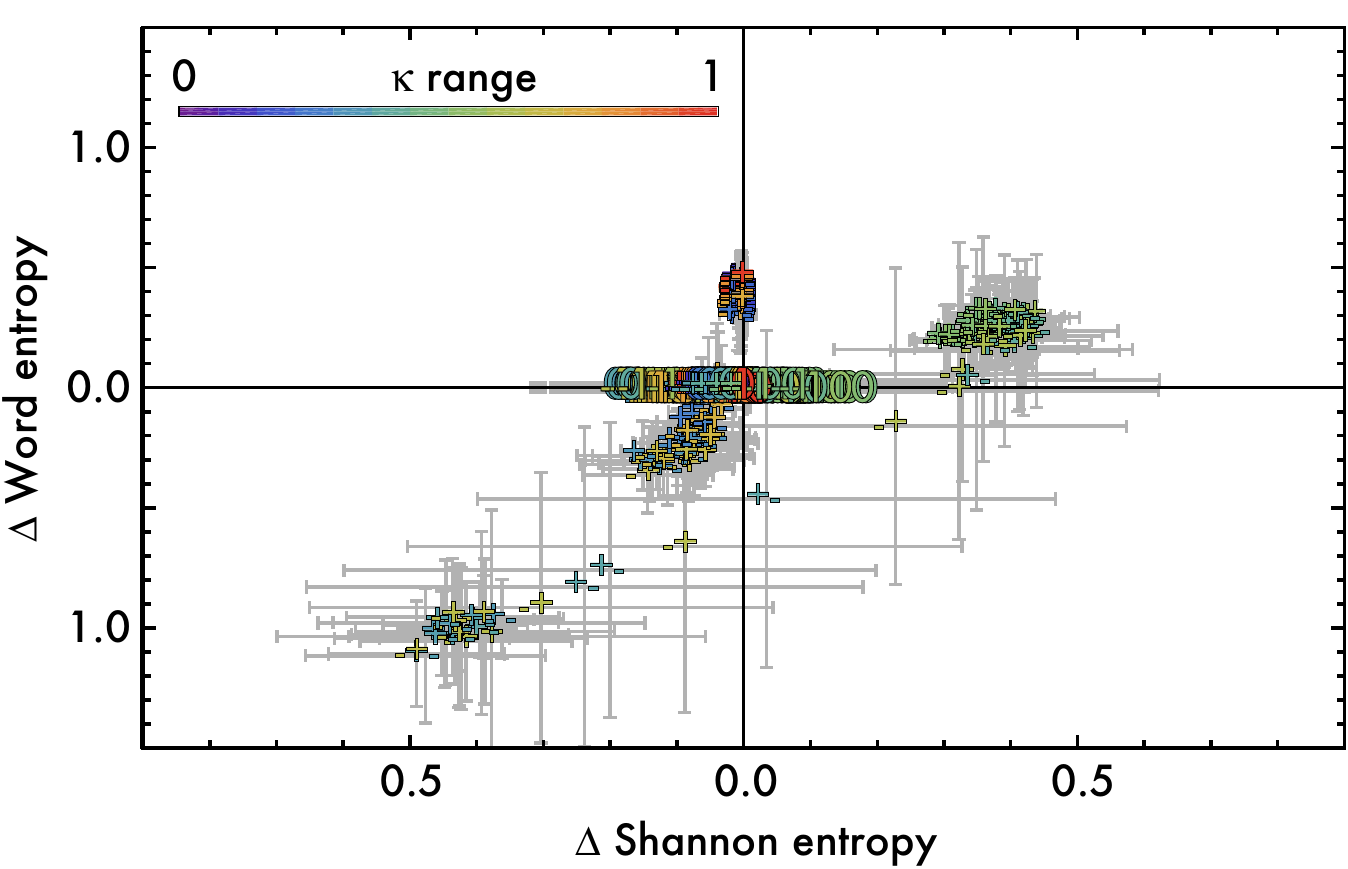}
\caption{Entropy signature difference plot for 25 scale-free graphs
  with 200 nodes and 400 links. Both positive and negative entropy
  differences ($\Delta E = E_\textrm{H} - E_\textrm{R}$) occur for the
  comparison of hierarchized vs. randomized graphs.}
  \label{sfGraphs}
\end{center}
\end{figure}

\textit{Complex networks}---The formalism of Eq.~(\ref{eq:3}) can be
transferred to networks of arbitrary topology. Compared to regular
graphs with global neighborhood size $d$, the case where $\rho_i =
\kappa$ will rarely occur in graphs with heterogeneous
connectivity. We thus simplify the set of possible rules with single
threshold: By setting $\alpha = \beta$ in Eq.~(\ref{eq:3}), the rule
space is condensed and a rule is now defined by the rule parameters
$(\alpha,\gamma)$ and the threshold parameter $\kappa$.

As a first exemplary application, we consider scale-free graphs,
generated by the incremental Barabási-Albert model \cite{barabasi99},
with a power law degree distribution, a property often found in
real-life networks \cite{albert02,newman03}. Due to their pivotal
topological property, the existence of hubs, scale-free graphs have
been used frequently as model graphs. They have also been used as a
starting point to investigate the relation of degree-degree
correlations in complex networks
\cite{maslov02_science,trusina04,weber08_europhysics-letters}. Here we
want to study how degree correlations in a scale-free graph affect its
ability to generate complex patterns.  We implement all resulting
rules on randomized, hierarchized and anti-hierarchized variants of
scale-free graphs with 200 nodes and 400 links. Hierarchization and
anti-hierarchization means the gradual randomization towards positive
and negative degree-degree correlations, respectively
\cite{trusina04}. For each graph type, we calculate the entropy
signatures, given by the Shannon entropy and word entropy of the
emerging patterns, for all rules. Figure \ref{sfGraphs} shows the
entropy signature difference plot, where $(S,W)$ of the randomized
graphs (R) has been subtracted from the entropy signature of the
hierarchized graphs (H), $\Delta E = E_\textrm{H} -
E_\textrm{R}$. Here, $E$ stands for $S$ and $W$ respectively. Most
rules are insensitive to degree-degree correlations.  Positive and
negative entropy signature differences occur preferentially for
$(\alpha,\beta) = (+,-)$ or $(\alpha,\beta) = (-,+)$ with $\kappa \in
[0.3,0.6]$. Notably, rule $(+,-)$ is a condensed form of the
topology-sensitive ECA rule 108, appearing in Figure
\ref{patterns}. For the anti-hierarchized graph with negative
degree-degree correlations, a similar picture emerges (data not
shown).

\begin{figure}[b]
\begin{center}
\includegraphics[width=0.8\columnwidth]{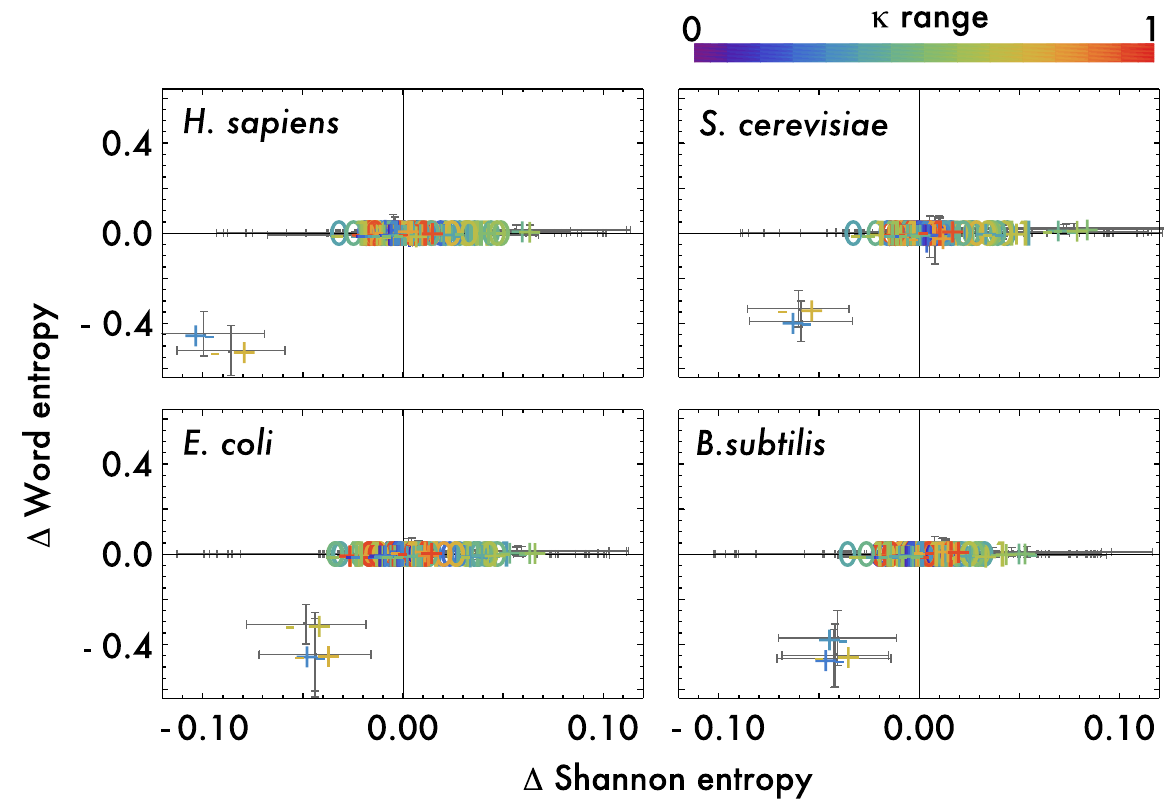}
\caption{Entropy signature difference plot for the metabolic
networks of \textit{H.~sapiens} $(N=625, M=779)$,
\textit{S.~cerevisiae} (448, 564), \textit{E.~coli} (563, 709), and
\textit{B.~subtilis} (501, 612).  $N$ denotes the number of nodes,
$M$ the number of links in the largest connected component of the
substrate graphs of the respective species. Notably, predominantly
negative differences $\Delta S$ and  $\Delta W$ occur.}
  \label{metNets}
\end{center}
\end{figure}

As a second example, we consider the topology of metabolic networks,
which abstracts the wiring architecture of the set of enzyme-catalyzed
reactions in a specific species. Substrate graphs, where nodes
represent metabolites and links represent a reaction between connected
substrates can be generated from genomic data \cite{ma03}. In
\cite{marr07_pre}, we recently studied the impact of the topology of
metabolic networks on a specific dynamics. There we implemented and
studied only a single rule, namely $(+,-)$ as a dynamic probe and
interpreted the enhanced regularizing capacity of real networks
compared to randomized null models as a possible topological
contribution to the reliable establishment of metabolic steady-states
and to the effective dampening of fluctuations. To show that the
results presented in \cite{marr07_pre} are valid over the whole range
of dynamics discussed in the present paper, we now implement all
possible rules of the form $(\alpha,\gamma)$ on substrate graphs and
analyze the entropy signature differences. Figure \ref{metNets} shows
the results for \textit{Homo sapiens}, \textit{Saccharomyces
  cerevisiae}, \textit{Escherichia coli}, and \textit{Bacillus
  subtilis}. We find that while the entropy signatures of most rules
do not discriminate between real and randomized topologies, a few
rules are topologically sensitive. These rules comply with
$(\alpha,\beta) =(+,-)$ or $(\alpha,\beta) = (-,+)$ while $\kappa \in
[0.35,0.55]$. For all these rules, the entropy signature of real
graphs is significantly smaller compared to the null model
topologies. This is also true for hierarchized and anti-hierarchized
null models, as well as for all other species investigated in
\cite{marr07_pre}. We believe that the application of dynamic probes
is a particularly helpful tool for studying dynamical constraints
imposed by topology.


\begin{table}[b]
  \centering
  \caption{Examples of symmetric outer-totalistic CA as defined in
    Eq.~(\ref{eq:2}), corresponding ECA and RBN rule numbers, as
    defined in \cite{wolfram83} and \cite{kauffman69}, respectively,
    and references where these rules have been discussed previously.  \label{tab:gca}}
  \begin{tabular}{c@{\hspace{3ex}}c@{\hspace{3ex}}c@{\hspace{3ex}}c}
    \hline\hline
    Outer-totalistic CA & ECA &  RBN & Ref.\\[.1cm]
      \hline
      $(0, 0, 0) \leftrightarrow (1, 1, 1)$ & $0 \leftrightarrow 255$
      & $1 \leftrightarrow 16$ & \\
      $(1, 0, 0) \leftrightarrow (1, 1, 0)$ & $5 \leftrightarrow 95$ &
      $2 \leftrightarrow 8$ & \cite{paul06_physical-review-e} \\
      $(1, -, +) \leftrightarrow (+, -, 0)$ & $22 \leftrightarrow 151$ & -& \cite{matache04,nagler05_pre} \\
      $(0, 1, 0) \leftrightarrow (1, 0, 1)$ & $90 \leftrightarrow 165$
      & $7 \leftrightarrow 10$ & \cite{nagler05_pre} \\
      $(+, +, -) \leftrightarrow (-, +, +)$ & $108 \leftrightarrow 201$ & - & \cite{marr06_physics-letters-a,marr05_physica-a,marr07_pre}\\
      $(+, 1, -) \leftrightarrow (-, 0, +)$ & $126 \leftrightarrow
      129$ & - & \cite{matache04,nagler05_pre}\\
      $(0, 0, 1) \leftrightarrow (0, 1, 1)$ & $160 \leftrightarrow
      250$ & $9 \leftrightarrow 15$ & \cite{nochomovitz06}\\
      $(0, 1, +) \leftrightarrow (+, 0, 1)$ & $164 \leftrightarrow
      218$ & - & \cite{nagler05_pre}\\
      $(0, +, 1)$ & $232$ & - &
      \cite{crutchfield95_proceedings-of-the-n,moreira04,amaral04,nochomovitz06}\\[.1cm]
    \hline\hline
  \end{tabular}
\end{table}

\textit{Discussion}---Our formalism can be used to describe
outer-totalistic CA and isotropic RBN rules in a common framework. It
allows the comprehensive discussion of previously introduced rule sets
on diverse topologies, like the selection of Boolean rules presented
in \cite{amaral04} or variations of the majority rule as used in
\cite{moreira04,nochomovitz06}. It moreover formalizes previous
attempts to generalize CA to graphs
\cite{osullivan01,darabos07_adv-complex-sys}, and is easily
extensible, e.g.~by introducing more than just one threshold parameter
or by using a larger state space. With the presented framework, the
often huge rule space of a discrete dynamical system can be
intuitively parametrized and systematically analyzed. The finding of
symmetric rules, for example, helps complementing specific CA
classes. The set of rules exhibiting power law spectra, as introduced
in \cite{nagler05_pre}, can thus be completed with the corresponding
symmetric counterparts. Also, many coarse-graining transitions between
CA rules, as presented in \cite{israeli06_phys.rev.e}, can be
immediately understood with a symmetry rational. Specific rule
generalizations, as discussed and analytically analyzed in
\cite{matache04} may be reconsidered from the more general perspective
provided in this letter. As a specific example, the application of ECA rule
22 to arbitrary graphs, stated as an open question in
\cite{matache04}, is straightforward with the presented
formalism. Limitations arise as soon as individual node
characteristics are to be taken into account. Still, the isotropic
subset of canalyzing Boolean rules, as discussed in
\cite{paul06_physical-review-e}, can be represented with our approach.
Table \ref{tab:gca} shows some examples of symmetric rules in our
formalism, the corresponding ECA and RBN rule number and references
where these rules have been previously applied.

An analysis of topologically sensitive rules with analytical tools as
developed in \cite{drossel05_phys.rev.lett.} or the recently
introduced basin entropy \cite{krawitz07_physical-review-lett} may
reveal state space changes associated with topological modifications.
Such analyses can elucidate dynamic properties also relevant for
regulatory dynamics of biological networks, which have been
successfully modeled with CA approaches
\cite{bornholdt00_proc-biol-sci, li04_pnas,davidich08_plos-one}. From this
perspective, our framework provides a means to comprehensively study
the sensitivity of a system to topological perturbations and
associated rule space modifications.


\begin{thebibliography}{33}
\expandafter\ifx\csname natexlab\endcsname\relax\def\natexlab#1{#1}\fi
\expandafter\ifx\csname bibnamefont\endcsname\relax
  \def\bibnamefont#1{#1}\fi
\expandafter\ifx\csname bibfnamefont\endcsname\relax
  \def\bibfnamefont#1{#1}\fi
\expandafter\ifx\csname citenamefont\endcsname\relax
  \def\citenamefont#1{#1}\fi
\expandafter\ifx\csname url\endcsname\relax
  \def\url#1{\texttt{#1}}\fi
\expandafter\ifx\csname urlprefix\endcsname\relax\def\urlprefix{URL }\fi
\providecommand{\bibinfo}[2]{#2}
\providecommand{\eprint}[2][]{\url{#2}}

\bibitem[{\citenamefont{von Neumann}(2001)}]{neumann63}
\bibinfo{author}{\bibfnamefont{J.}~\bibnamefont{von Neumann}}, in
  \emph{\bibinfo{booktitle}{J.~von Neumann, Collected Works}}, edited by
  \bibinfo{editor}{\bibfnamefont{A.~H.} \bibnamefont{Taub}}
  (\bibinfo{publisher}{Macmillan}, \bibinfo{address}{New York},
  \bibinfo{year}{2001}), vol.~\bibinfo{volume}{5}, p. \bibinfo{pages}{288}.

\bibitem[{\citenamefont{Wolfram}(1983)}]{wolfram83}
\bibinfo{author}{\bibfnamefont{S.}~\bibnamefont{Wolfram}},
  \bibinfo{journal}{Rev. Mod. Phys.} \textbf{\bibinfo{volume}{55}},
  \bibinfo{pages}{601} (\bibinfo{year}{1983}).

\bibitem[{\citenamefont{Deutsch and Dormann}(2005)}]{deutsch05}
\bibinfo{author}{\bibfnamefont{A.}~\bibnamefont{Deutsch}} \bibnamefont{and}
  \bibinfo{author}{\bibfnamefont{S.}~\bibnamefont{Dormann}},
  \emph{\bibinfo{title}{Cellular Automaton Modeling and Biological Pattern
  Formation}} (\bibinfo{publisher}{Birkhäuser}, \bibinfo{address}{Boston},
  \bibinfo{year}{2005}).

\bibitem[{\citenamefont{Kauffman}(1969)}]{kauffman69}
\bibinfo{author}{\bibfnamefont{S.~A.} \bibnamefont{Kauffman}},
  \bibinfo{journal}{J. Theor. Biol.} \textbf{\bibinfo{volume}{22}},
  \bibinfo{pages}{437} (\bibinfo{year}{1969}).

\bibitem[{\citenamefont{Crutchfield and
  Mitchell}(1995)}]{crutchfield95_proceedings-of-the-n}
\bibinfo{author}{\bibfnamefont{J.}~\bibnamefont{Crutchfield}} \bibnamefont{and}
  \bibinfo{author}{\bibfnamefont{M.}~\bibnamefont{Mitchell}},
  \bibinfo{journal}{Proc. Natl. Acad. Sci. USA} \textbf{\bibinfo{volume}{92}},
  \bibinfo{pages}{10742} (\bibinfo{year}{1995}).

\bibitem[{\citenamefont{Moreira et~al.}(2004)\citenamefont{Moreira, Mathur,
  Diermeier, and Amaral}}]{moreira04}
\bibinfo{author}{\bibfnamefont{A.~A.} \bibnamefont{Moreira}},
  \bibinfo{author}{\bibfnamefont{A.}~\bibnamefont{Mathur}},
  \bibinfo{author}{\bibfnamefont{D.}~\bibnamefont{Diermeier}},
  \bibnamefont{and} \bibinfo{author}{\bibfnamefont{L.~A.~N.}
  \bibnamefont{Amaral}}, \bibinfo{journal}{Proc. Natl. Acad. Sci. USA}
  \textbf{\bibinfo{volume}{101}}, \bibinfo{pages}{12085}
  (\bibinfo{year}{2004}).

\bibitem[{\citenamefont{Amaral et~al.}(2004)\citenamefont{Amaral,
  D{\'\i}az-Guilera, Moreira, Goldberger, and Lipsitz}}]{amaral04}
\bibinfo{author}{\bibfnamefont{L.~A.~N.} \bibnamefont{Amaral}},
  \bibinfo{author}{\bibfnamefont{A.}~\bibnamefont{D{\'\i}az-Guilera}},
  \bibinfo{author}{\bibfnamefont{A.~A.} \bibnamefont{Moreira}},
  \bibinfo{author}{\bibfnamefont{A.~L.} \bibnamefont{Goldberger}},
  \bibnamefont{and} \bibinfo{author}{\bibfnamefont{L.~A.}
  \bibnamefont{Lipsitz}}, \bibinfo{journal}{Proc. Natl. Acad. Sci. USA}
  \textbf{\bibinfo{volume}{101}}, \bibinfo{pages}{15551}
  (\bibinfo{year}{2004}).

\bibitem[{\citenamefont{Nochomovitz and Li}(2006)}]{nochomovitz06}
\bibinfo{author}{\bibfnamefont{Y.~D.} \bibnamefont{Nochomovitz}}
  \bibnamefont{and} \bibinfo{author}{\bibfnamefont{H.}~\bibnamefont{Li}},
  \bibinfo{journal}{Proc. Natl. Acad. Sci. USA} \textbf{\bibinfo{volume}{103}},
  \bibinfo{pages}{4180} (\bibinfo{year}{2006}).

\bibitem[{\citenamefont{Wolfram}(1984)}]{wolfram84}
\bibinfo{author}{\bibfnamefont{S.}~\bibnamefont{Wolfram}},
  \bibinfo{journal}{Physica D} \textbf{\bibinfo{volume}{10}},
  \bibinfo{pages}{1} (\bibinfo{year}{1984}).

\bibitem[{\citenamefont{Marr and H{\"u}tt}(2006)}]{marr06_physics-letters-a}
\bibinfo{author}{\bibfnamefont{C.}~\bibnamefont{Marr}} \bibnamefont{and}
  \bibinfo{author}{\bibfnamefont{M.-T.} \bibnamefont{H{\"u}tt}},
  \bibinfo{journal}{Phys. Lett. A} \textbf{\bibinfo{volume}{349}},
  \bibinfo{pages}{302} (\bibinfo{year}{2006}).

\bibitem[{\citenamefont{Marr and H{\"u}tt}(2005)}]{marr05_physica-a}
\bibinfo{author}{\bibfnamefont{C.}~\bibnamefont{Marr}} \bibnamefont{and}
  \bibinfo{author}{\bibfnamefont{M.-T.} \bibnamefont{H{\"u}tt}},
  \bibinfo{journal}{Physica A} \textbf{\bibinfo{volume}{354}},
  \bibinfo{pages}{641} (\bibinfo{year}{2005}).

\bibitem[{\citenamefont{Marr et~al.}(2007)\citenamefont{Marr, M{\"u}ller-Linow,
  and H{\"u}tt}}]{marr07_pre}
\bibinfo{author}{\bibfnamefont{C.}~\bibnamefont{Marr}},
  \bibinfo{author}{\bibfnamefont{M.}~\bibnamefont{M{\"u}ller-Linow}},
  \bibnamefont{and} \bibinfo{author}{\bibfnamefont{M.-T.}
  \bibnamefont{H{\"u}tt}}, \bibinfo{journal}{Phys. Rev. E}
  \textbf{\bibinfo{volume}{75}}, \bibinfo{pages}{041917}
  (\bibinfo{year}{2007}).

\bibitem[{\citenamefont{Trusina et~al.}(2004)\citenamefont{Trusina, Maslov,
  Minnhagen, and Sneppen}}]{trusina04}
\bibinfo{author}{\bibfnamefont{A.}~\bibnamefont{Trusina}},
  \bibinfo{author}{\bibfnamefont{S.}~\bibnamefont{Maslov}},
  \bibinfo{author}{\bibfnamefont{P.}~\bibnamefont{Minnhagen}},
  \bibnamefont{and} \bibinfo{author}{\bibfnamefont{K.}~\bibnamefont{Sneppen}},
  \bibinfo{journal}{Phys. Rev. Lett.} \textbf{\bibinfo{volume}{92}},
  \bibinfo{pages}{178702} (\bibinfo{year}{2004}).

\bibitem[{\citenamefont{Wormald}(1999)}]{wormald99_randreggraphs}
\bibinfo{author}{\bibfnamefont{N.~C.} \bibnamefont{Wormald}}, in
  \emph{\bibinfo{booktitle}{Surveys in Combinatorics}}
  (\bibinfo{publisher}{Cambridge University Press}, \bibinfo{year}{1999}), pp.
  \bibinfo{pages}{239--298}.

\bibitem[{\citenamefont{Erd\H{o}s and R{\'e}nyi}(1959)}]{erdos59}
\bibinfo{author}{\bibfnamefont{P.}~\bibnamefont{Erd\H{o}s}} \bibnamefont{and}
  \bibinfo{author}{\bibfnamefont{A.}~\bibnamefont{R{\'e}nyi}},
  \bibinfo{journal}{Publ. Math. (Debrecen)} \textbf{\bibinfo{volume}{6}},
  \bibinfo{pages}{290} (\bibinfo{year}{1959}).

\bibitem[{\citenamefont{Peng et~al.}(1994)\citenamefont{Peng, Buldyrev, Havlin,
  Simons, Stanley, and Goldberger}}]{peng94}
\bibinfo{author}{\bibfnamefont{C.-K.} \bibnamefont{Peng}},
  \bibinfo{author}{\bibfnamefont{S.~V.} \bibnamefont{Buldyrev}},
  \bibinfo{author}{\bibfnamefont{S.}~\bibnamefont{Havlin}},
  \bibinfo{author}{\bibfnamefont{M.}~\bibnamefont{Simons}},
  \bibinfo{author}{\bibfnamefont{H.~E.} \bibnamefont{Stanley}},
  \bibnamefont{and} \bibinfo{author}{\bibfnamefont{A.~L.}
  \bibnamefont{Goldberger}}, \bibinfo{journal}{Phys. Rev. E}
  \textbf{\bibinfo{volume}{49}}, \bibinfo{pages}{1685} (\bibinfo{year}{1994}).

\bibitem[{\citenamefont{Barab{\'a}si and Albert}(1999)}]{barabasi99}
\bibinfo{author}{\bibfnamefont{A.-L.} \bibnamefont{Barab{\'a}si}}
  \bibnamefont{and} \bibinfo{author}{\bibfnamefont{R.}~\bibnamefont{Albert}},
  \bibinfo{journal}{Science} \textbf{\bibinfo{volume}{286}},
  \bibinfo{pages}{509} (\bibinfo{year}{1999}).

\bibitem[{\citenamefont{Albert and Barab{\'a}si}(2002)}]{albert02}
\bibinfo{author}{\bibfnamefont{R.}~\bibnamefont{Albert}} \bibnamefont{and}
  \bibinfo{author}{\bibfnamefont{A.-L.} \bibnamefont{Barab{\'a}si}},
  \bibinfo{journal}{Rev. Mod. Phys.} \textbf{\bibinfo{volume}{74}},
  \bibinfo{pages}{47} (\bibinfo{year}{2002}).

\bibitem[{\citenamefont{Newman}(2003)}]{newman03}
\bibinfo{author}{\bibfnamefont{M.~E.~J.} \bibnamefont{Newman}},
  \bibinfo{journal}{SIAM Rev.} \textbf{\bibinfo{volume}{45}},
  \bibinfo{pages}{167} (\bibinfo{year}{2003}).

\bibitem[{\citenamefont{Maslov and Sneppen}(2002)}]{maslov02_science}
\bibinfo{author}{\bibfnamefont{S.}~\bibnamefont{Maslov}} \bibnamefont{and}
  \bibinfo{author}{\bibfnamefont{K.}~\bibnamefont{Sneppen}},
  \bibinfo{journal}{Science} \textbf{\bibinfo{volume}{296}},
  \bibinfo{pages}{910} (\bibinfo{year}{2002}).

\bibitem[{\citenamefont{Weber et~al.}(2008)\citenamefont{Weber, H{\"u}tt, and
  Porto}}]{weber08_europhysics-letters}
\bibinfo{author}{\bibfnamefont{S.}~\bibnamefont{Weber}},
  \bibinfo{author}{\bibfnamefont{M.}~\bibnamefont{H{\"u}tt}}, \bibnamefont{and}
  \bibinfo{author}{\bibfnamefont{M.}~\bibnamefont{Porto}},
  \bibinfo{journal}{Europhysics Letters} \textbf{\bibinfo{volume}{82}},
  \bibinfo{pages}{28003} (\bibinfo{year}{2008}).

\bibitem[{\citenamefont{Ma and Zeng}(2003)}]{ma03}
\bibinfo{author}{\bibfnamefont{H.}~\bibnamefont{Ma}} \bibnamefont{and}
  \bibinfo{author}{\bibfnamefont{A.-P.} \bibnamefont{Zeng}},
  \bibinfo{journal}{Bioinformatics} \textbf{\bibinfo{volume}{19}},
  \bibinfo{pages}{270} (\bibinfo{year}{2003}).

\bibitem[{\citenamefont{Paul et~al.}(2006)\citenamefont{Paul, Kaufman, and
  Drossel}}]{paul06_physical-review-e}
\bibinfo{author}{\bibfnamefont{U.}~\bibnamefont{Paul}},
  \bibinfo{author}{\bibfnamefont{V.}~\bibnamefont{Kaufman}}, \bibnamefont{and}
  \bibinfo{author}{\bibfnamefont{B.}~\bibnamefont{Drossel}},
  \bibinfo{journal}{Phys. Rev. E} \textbf{\bibinfo{volume}{73}},
  \bibinfo{pages}{26118} (\bibinfo{year}{2006}).

\bibitem[{\citenamefont{Nagler and Claussen}(2005)}]{nagler05_pre}
\bibinfo{author}{\bibfnamefont{J.}~\bibnamefont{Nagler}} \bibnamefont{and}
  \bibinfo{author}{\bibfnamefont{J.~C.} \bibnamefont{Claussen}},
  \bibinfo{journal}{Phys. Rev. E} \textbf{\bibinfo{volume}{71}},
  \bibinfo{pages}{067103} (\bibinfo{year}{2005}).

\bibitem[{\citenamefont{Matache and Heidel}(2004)}]{matache04}
\bibinfo{author}{\bibfnamefont{M.~T.} \bibnamefont{Matache}} \bibnamefont{and}
  \bibinfo{author}{\bibfnamefont{J.}~\bibnamefont{Heidel}},
  \bibinfo{journal}{Phys. Rev. E} \textbf{\bibinfo{volume}{69}},
  \bibinfo{pages}{056214} (\bibinfo{year}{2004}).

\bibitem[{\citenamefont{O'Sullivan}(2001)}]{osullivan01}
\bibinfo{author}{\bibfnamefont{D.}~\bibnamefont{O'Sullivan}},
  \bibinfo{journal}{Environment and Planning B} \textbf{\bibinfo{volume}{28}},
  \bibinfo{pages}{687} (\bibinfo{year}{2001}).

\bibitem[{\citenamefont{Darabos et~al.}(2007)\citenamefont{Darabos, Giacobini,
  and Tomassini}}]{darabos07_adv-complex-sys}
\bibinfo{author}{\bibfnamefont{C.}~\bibnamefont{Darabos}},
  \bibinfo{author}{\bibfnamefont{M.}~\bibnamefont{Giacobini}},
  \bibnamefont{and}
  \bibinfo{author}{\bibfnamefont{M.}~\bibnamefont{Tomassini}},
  \bibinfo{journal}{Advances in Complex Systems} \textbf{\bibinfo{volume}{10}},
  \bibinfo{pages}{85} (\bibinfo{year}{2007}).

\bibitem[{\citenamefont{Israeli and Goldenfeld}(2006)}]{israeli06_phys.rev.e}
\bibinfo{author}{\bibfnamefont{N.}~\bibnamefont{Israeli}} \bibnamefont{and}
  \bibinfo{author}{\bibfnamefont{N.}~\bibnamefont{Goldenfeld}},
  \bibinfo{journal}{Phys. Rev. E} \textbf{\bibinfo{volume}{73}},
  \bibinfo{pages}{026203} (\bibinfo{year}{2006}).

\bibitem[{\citenamefont{Drossel et~al.}(2005)\citenamefont{Drossel, Mihaljev,
  and Greil}}]{drossel05_phys.rev.lett.}
\bibinfo{author}{\bibfnamefont{B.}~\bibnamefont{Drossel}},
  \bibinfo{author}{\bibfnamefont{T.}~\bibnamefont{Mihaljev}}, \bibnamefont{and}
  \bibinfo{author}{\bibfnamefont{F.}~\bibnamefont{Greil}},
  \bibinfo{journal}{Phys. Rev. Lett.} \textbf{\bibinfo{volume}{94}},
  \bibinfo{pages}{88701} (\bibinfo{year}{2005}).

\bibitem[{\citenamefont{Krawitz and
  Shmulevich}(2007)}]{krawitz07_physical-review-lett}
\bibinfo{author}{\bibfnamefont{P.}~\bibnamefont{Krawitz}} \bibnamefont{and}
  \bibinfo{author}{\bibfnamefont{I.}~\bibnamefont{Shmulevich}},
  \bibinfo{journal}{Phys. Rev. Lett.} \textbf{\bibinfo{volume}{98}},
  \bibinfo{eid}{158701} (\bibinfo{year}{2007}).

\bibitem[{\citenamefont{Bornholdt and
  Sneppen}(2000)}]{bornholdt00_proc-biol-sci}
\bibinfo{author}{\bibfnamefont{S.}~\bibnamefont{Bornholdt}} \bibnamefont{and}
  \bibinfo{author}{\bibfnamefont{K.}~\bibnamefont{Sneppen}},
  \bibinfo{journal}{Proc. R. Soc. Lond. B} \textbf{\bibinfo{volume}{267}},
  \bibinfo{pages}{2281} (\bibinfo{year}{2000}).

\bibitem[{\citenamefont{Li et~al.}(2004)\citenamefont{Li, Long, Lu, Ouyang, and
  Tang}}]{li04_pnas}
\bibinfo{author}{\bibfnamefont{F.}~\bibnamefont{Li}},
  \bibinfo{author}{\bibfnamefont{T.}~\bibnamefont{Long}},
  \bibinfo{author}{\bibfnamefont{Y.}~\bibnamefont{Lu}},
  \bibinfo{author}{\bibfnamefont{Q.}~\bibnamefont{Ouyang}}, \bibnamefont{and}
  \bibinfo{author}{\bibfnamefont{C.}~\bibnamefont{Tang}},
  \bibinfo{journal}{Proc. Natl. Acad. Sci. USA} \textbf{\bibinfo{volume}{101}},
  \bibinfo{pages}{4781} (\bibinfo{year}{2004}).

\bibitem[{\citenamefont{Davidich and Bornholdt}(2008)}]{davidich08_plos-one}
\bibinfo{author}{\bibfnamefont{M.}~\bibnamefont{Davidich}} \bibnamefont{and}
  \bibinfo{author}{\bibfnamefont{S.}~\bibnamefont{Bornholdt}},
  \bibinfo{journal}{PLoS ONE} \textbf{\bibinfo{volume}{3}}
  (\bibinfo{year}{2008}).

\end{thebibliography}

\end{document}